\documentclass[apjl]{emulateapj}
\usepackage{hyperref}
\usepackage{color}
\hypersetup{colorlinks,
 citecolor=blue,
 linkcolor=blue}
\usepackage{natbib}

\usepackage{ifthen} 
\usepackage[table]{xcolor}

\def\aj{AJ}
\def\apj{ApJ}
\def\apjl{ApJ}
\def\apjs{ApJS}
\def\aap{A\&A}
\def\aapr{A\&AR}
\def\mnras{MNRAS}
\def\nat{Nature}
\def\pasp{PASP}%
\def\araa{ARA\&A}

\definecolor{lr}{rgb}{1,0.8,0.8} 
\definecolor{lg}{rgb}{0.8,1,0.8} 
\definecolor{lb}{rgb}{0.8,0.8,1} 
\definecolor{lk}{rgb}{0.8,0.8,0.8} 

\newcommand{\tclr}[1]{{\color{red}{#1}}}

\slugcomment{Draft Version \today}
\shorttitle{Shells in Virgo early-type dwarf galaxies}
\shortauthors{Paudel et al.}
\begin{document}
\title{THE NEXT GENERATION VIRGO CLUSTER SURVEY. XIV. Shell feature early-type dwarf  galaxies in the Virgo cluster\altaffilmark{*}}
\altaffiltext{*}{Based on observations obtained with MegaPrime/MegaCam, a joint project of CFHT and CEA/DAPNIA, at the Canada-France-Hawaii Telescope (CFHT) which is operated by the National Research Council (NRC) of Canada, the Institut National des Sciences de l'Univers of the Centre National de la Recherche Scientifique of France and the University of Hawaii.}

\author{
Sanjaya Paudel\altaffilmark{1},
Rory Smith \altaffilmark{2}, 
Pierre-Alain Duc \altaffilmark{3},
Patrick C\^ot\'e\altaffilmark{4},
Jean-Charles Cuillandre\altaffilmark{3},
Laura Ferrarese\altaffilmark{4},
John P. Blakeslee\altaffilmark{4},
Alessandro Boselli\altaffilmark{5},
Michele Cantiello\altaffilmark{6}
S.D.J. Gwyn\altaffilmark{4},
Puragra Guhathakurta\altaffilmark{7},
Simona Mei\altaffilmark{8,9,10},
J. Christopher Mihos\altaffilmark{11},
Eric W. Peng\altaffilmark{12,13},
Mathieu Powalka\altaffilmark{14},
R\'uben S\'anchez-Janssen\altaffilmark{4},
Elisa Toloba\altaffilmark{15} \& 
Hongxin Zhang\altaffilmark{12,16}
}

\affil{$^1$ Korea Astronomy and Space Science Institute, Daejon 305-348, Republic of Korea; sanjpaudel@gmail.com}
\affil{$^2$Yonsei University, Graduate School of Earth System Sciences-Astronomy-Atmospheric Sciences, Seoul 120-749, Republic of Korea}
\affil{$^3$ Laboratoire AIM Paris-Saclay, CEA/IRFU/SAp, 91191 Gif-sur-Yvette Cedex, France}
\affil{$^4$ NRC Herzberg Astronomy and Astrophysics, 5071 West Saanich Road, Victoria, BC V9E 2E7, Canada}
\affil{$^5$Aix Marseille Universit\'e, CNRS, LAM (Laboratoire d'Astrophysique de Marseille) UMR 7326, 13388, Marseille, France}
\affil{$^6$INAF Osservatorio Astronomico di Teramo, via Maggini snc, 64100, Teramo, Italy}

\affil{$^7$UCO/Lick Observatory, University of California, Santa Cruz, 1156 High Street, Santa Cruz, CA 95064, USA}
\affil{$^{8}$ GEPI, Observatoire de Paris, CNRS, Universit\'e Paris Diderot, Paris Sciences et Lettres (PSL), 61, Avenue de l'Observatoire 75014, Paris France}
\affil{$^{9}$ Universit\'{e} Paris Denis Diderot, Universit\'e Paris Sorbonne Cit\'e, 75205 Paris Cedex13, France}
\affil{$^{10}$ Jet Propulsion Laboratory, California Institute of Technology, Pasadena, CA 91125, USA}
\affil{$^{11}$Department of Astronomy, Case Western Reserve University, Cleveland, OH 41106, USA}
\affil{$^{12}$Department of Astronomy, Peking University, Beijing 100871, China}
\affil{$^{13}$Kavli Institute for Astronomy and Astrophysics, Peking University, Beijing 100871, China}
\affil{$^{14}$Observatoire Astronomique de Strasbourg, Universit\'e de Strasbourg, CNRS, UMR 7550, 11 rue de l'Universit\'e, F-67000 Strasbourg, France}
\affil{$^{15}$Department of Physics, Texas Tech University, Box 41051, Lubbock, TX}
\affil{$^{16}$Instituto de Astrof\'{\i}sica, Pontificia Universidad Cat\'olica de Chile, 7820436 Macul, Santiago, Chile}

\begin{abstract}
The Next Generation Virgo Cluster Survey is a deep (with a $2\sigma$ detection limit $\mu_g$ = 29~mag~arcsec$^{-2}$ in the $g-$band) optical panchromatic survey targeting the Virgo cluster from its core to virial radius, for a total areal coverage of 104 square degrees. As such, the survey is well suited for the study of galaxies' outskirts, haloes and low surface brightness features that arise from dynamical interactions within the cluster environment. We report the discovery of extremely faint ($\mu_g$  $>$ 25 mag arcsec$^{-2}$) shells in three Virgo cluster early-type dwarf galaxies, VCC~1361, VCC~1447 and VCC~1668. Among them, VCC~1447 has an absolute magnitude  M$_{g}$ = -11.71 mag and is {\it the least massive galaxy with a shell system discovered to date}. We present a detailed study of these low surface brightness features. We detect between three and four shells in each of our galaxies. Within the uncertainties, we find no evidence of a color difference between the galaxy main body and shell features. The observed arcs of the shells are located upto several effective radii of the galaxies. We further explore the origin of these low surface brightness features with the help of idealized numerical simulations. We find that a near equal mass merger is best able to reproduce the main properties of the shells, including their quite symmetric appearance and their alignment along the major axis of the galaxy. The simulations provide support for a formation scenario in which a recent merger, between two near-equal mass, gas-free dwarf galaxies forms the observed shell systems.
\end{abstract}


\section{Introduction}

Fine structures in deep imaging of galaxies are often used as supporting evidence that galaxies are evolving within the hierarchical merging scenario. Features such as shells, stellar streams, filaments and tidal tails are commonly found in deep imaging surveys of massive galaxies  \citep{Schweizer82,Turnbull99,Wilkinson00,Cooper11,Duc15}. However, when it comes to lower mass galaxies, such features are found much less frequently. In the hierarchical cosmology scenario, low mass galaxies are the first objects to form. In fact they are believed to be the building blocks of all more massive galaxies, who grow from these first blocks through a process of merging. Therefore many of today's low mass early-type galaxies, so-called dwarf spheroidal galaxies (dSph), are believed to be primordial galaxies -- the left over building blocks of galaxy formation events \citep{Grebel03,Ricotti05}. 

Faint shell-like features were first reported by \cite{Arp66}. Since then, they have been studied with great interest, both observationally and theoretically. A variety of models have been proposed for the formation of shells \citep{Quinn84,Thomson91,Hernquist92}. It is generally accepted that the shells are the results of phase wrapping or spatial wrapping \citep{Quinn84}. Early studies mainly discussed shell formation in the context of minor mergers, where the shells are the stellar remnant of an accreted secondary galaxy \citep{Quinn84,Dupraz86,Hernquist88,Hernquist89}. However, \cite{Hernquist92} showed that major mergers can also produce shell systems that are similar to the observed ones. Therefore, it is not clear whether shell formation happens preferentially in major or minor mergers, or equally in both. Either way, the recent literature strongly supports the picture that shells form during mergers. Therefore, the presence of shell features has a clear interpretation. Other typical low surface brightness features such as stellar streams, and filaments, are likely the result of tidal interactions. The difference between shells and these other features is that the shells are only present after the merger has occurred, whereas streams may be formed before the merger, during a tidal stripping phase \cite[e.g][]{Paudel13}.

A classification scheme of shell system was proposed by \cite{Prieur90}, where interleaving shells that are aligned along the major axis of the galaxy can be classified as type-I systems. Type-II systems have randomly distributed shells all around the galaxy. NGC~7600 and NGC~474 are well known examples of type-I and type-II shell system, respectively \citep{Cooper11,Duc15}. Type-III systems are those which can not be classified as type-I or type-II. \cite{Hernquist92} showed that type-I shell system can be formed during equal mass mergers. Good examples of Type-I systems include NGC 3923 and NGC~7600. The former is extensively reviewed by \cite{Carter98} \citep[also see][]{Bilek16}, and a visual replica of the latter has been reproduced via mergers, by \cite{Cooper11}, using cosmological simulations.

With the discovery of a shell-like feature in the Fornax dwarf galaxy, there has been a plethora of observational and theoretical studies, interpreting and discussing the origin of this structure in the recent literature \citep{Coleman04,Coleman05,Olszewski06,Coleman08,Bate15,delPino13}. The most likely interpretation is that Fornax has formed through a minor merger of two dwarf galaxies \citep{Yozin12}. Until now, Fornax was the only galaxy of mass M$_{*}$  $<$ 1$\times$10$^{8}$M$_{\sun}$ known to possess shell features. 
However a number of authors have disputed the true nature of the observed shell-like feature \citep[e.g.][]{Penarrubia09,Bate15}. As it is nearby, the stars can be resolved individually, enabling one to carry out a detailed kinematic and stellar population study.

Low mass early-type galaxies, generally referred to as dwarf Ellipticals (dEs), dominate the cluster and group environment by number. They are very rare in isolation or in the field environment \citep{Binggeli85,Geha12}. Such an extreme form of the so-called morphology-density correlation favors the idea that the local environment plays a significant role in the evolution of these galaxies \citep{Boselli14b}. Moreover, it is commonly believed that, by having a shallow potential well, low-mass galaxies should be more significantly influenced by their surrounding environment, and less so by merging events. Nevertheless, mounting evidence suggests that galaxy-galaxy mergers in low mass galaxies might not be as rare as was previously thought, and there is much speculation on the importance of mergers for the observed structural, kinematic, and stellar population properties of many dwarf galaxies \citep[e.g][]{Geha05,Graham12,Amorisco14,Paudel14,Toloba14,Pak16}. Very recently, direct observational evidence for mergers between dwarf galaxies has been discovered \citep[e.g.][]{Delgado12,Penny12,Rich12,Johnson13,Nidever13,Crnojevic14,Paudel15,Stierwalt15}. Furthermore, for the first time, the latest numerical simulations are able to reach the required resolution in order to study the effects of dwarf-dwarf mergers in a cosmological scenario \citep{Klimentowski10,Kazantzidis11,Deason14,Llambay16}, and these demonstrate that dwarf-dwarf mergers are a viable mechanism to form dwarf spheroidals in the Local Group environment  \citep{Kazantzidis11}.

In this study, we report the discovery of shell systems in three similar mass ($\approx$5$\times$10$^{7}$ M$_{\sun}$) dwarf galaxies, in the Virgo cluster, using images from the deep Next Generation Virgo cluster Survey \citep[NGVS,][]{Ferrarese12}. The observed shell features are much more pronounced than that of the Fornax dwarf galaxy. Among them, VCC~1447, the faintest in our sample, is nearly one magnitude fainter than Fornax, making it the faintest galaxy known to present shell features to date. We present a detailed study of the morphology and stellar population properties of these low surface brightness features and structural properties of the host dwarf galaxies. We compare these observations to idealised numerical simulations of merging dwarf galaxies, and find that their properties are best reproduced by near equal mass mergers between two early-type dwarfs. The detection of these features may offer a unique opportunity to study the ongoing, but dramatic, evolution of dwarf galaxies, a process that is often assumed to have finished long ago, in the cluster environment \citep{Behroozi2014}.

\section{Detection and simulations of shells in dwarf galaxies}

\subsection{Observed sample}

\begin{figure}
\includegraphics[width=7.5cm]{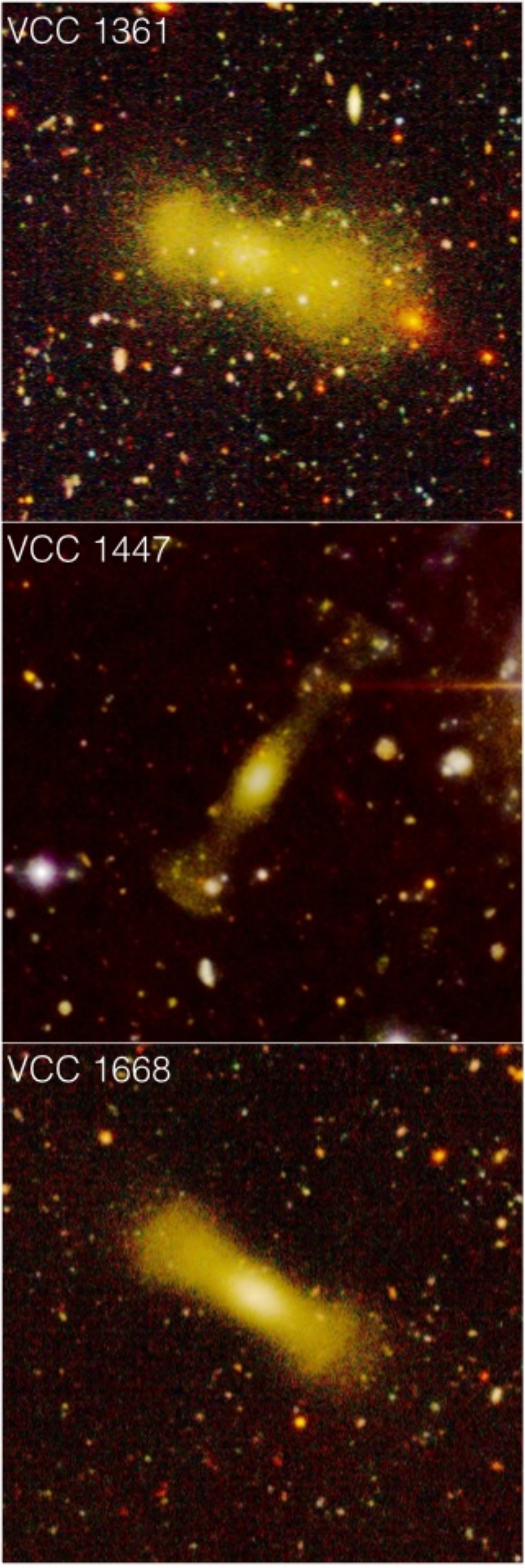}
\caption{Composite true color images made from NGVS $g-, i-, z-$band images. The field of view is  2\arcmin$\times$\,2$\arcmin$.  }
\label{vshel}
\end{figure}

The NGVS provides the deepest large scale map of the Virgo cluster to date, with a limiting magnitude for extended low-surface brightness object of $\sim$29 mag arcsec$^{-2}$ \citep{Ferrarese12,Ferrarese16}. It covers a total of 104 sq.deg area of the Virgo cluster, and is centered on the two sub-groups surrounding the massive early-type galaxies M87, and M49. 

NGVS images were obtained with the MegaCam imager, mounted on the Canada France Telescope (CHFT). The details of the observation strategy, and reduction procedure, can be found in \cite{Ferrarese12}. For this study, we mainly used the images obtained in  the $g'$ and $i'$ bands\footnote{Noted as $g$ and $i$ in the rest of the paper}. Since the $g$-band image is the most sensitive and the cleanest (i.e. it contains the least number of artifacts such as CCD imprints and reflection halos as discussed below), it was preferentially used for our photometric analysis.

One of the main aims of the NGVS project is to detect and catalog low-surface brightness features, such as stellar streams, filaments and shells around Virgo cluster member galaxies. As such, we perform a systematic search of these objects by visually inspecting all galaxies\footnote{We use the Virgo Cluster Catalogue \citep[VCC][]{Binggeli85} to select the parent sample}. A full catalogue of these low surface brightness features, with their morphological classifications, will be presented in a future paper. In this work, we present a detailed analysis of a sample of low mass (M$_{*}$ $<$ 10$^{8}$ M$_{\sun}$) early-type galaxies where shell features are prominent.

Traditionally, in larger galaxies, the search for shell features has occurred predominantly in early type galaxies. We select three early-type dwarf galaxies, VCC 1361,VCC 1447 and VCC 1668 for this study, in which prominent shell features are visible in the NGVS imaging. However, we note that this is not a complete sample of dwarf galaxies (M$_{B}$ $>$ $-$18 mag) with shell in the Virgo cluster. We have found that there are many other such galaxies that present shell-like features. However, they are typically more massive than the galaxies in this sample.

Carefully visually inspecting the NGVS images of VCC dwarf galaxies covered by the NGVS survey area, we find that 98 dwarf galaxies contains low surface brightness features that could be classified as shells. Among them, 64 are dE/dS0, there are 5 BCDs, and the rest are a mixture of types, such as dIrr, dPec, dSp etc. Generally, an observed break in the surface brightness map, with an arc shape, is referred to as a shell. In the case of dwarf galaxies, they are not as pronounced as in the case of giant galaxies, and are more ambiguous. For example, the shell feature in the Fornax dwarf spheroidal galaxy (See Colman et al 2004, Figure 1) is merely a small, over-dense region. However, such features might not have formed exclusively via mergers. There is much debate on the origin of the shell in the Fornax dwarf -- whether it is a dissolved star cluster, or a signature of a past merger. A break in surface brightness can also occur as a result of disk features, such as spiral arms, or in the case of a tidally truncated disk. Therefore it
is difficult to confirm that all 98 dwarf galaxies showing shell-like features are indeed definitive shells.

For this study, we have selected among the candidate list those that are non star-forming dwarf galaxies, of stellar mass similar to Fornax dwarf galaxy, and presenting indisputable shell features. The shells appear as external features. Since they are not embedded in the main body it is easy to confirm their presence. No extended/stretched spiral arm/bar or S-shape features are observed around these galaxies, i.e features which are produced when dwarf galaxies tidally interact with a nearby giant companion \citep[e.g.][]{Paudel13,Paudel14b}. This suggests that the shells are unlikely to be caused by a significant, ongoing tidal interaction with a nearby galaxy. We study these galaxies with the aim to explore the importance of mergers in the formation and evolution of dEs, which are the dominant galaxy population in the Virgo cluster. The depth of the NGVS images, and the image quality, permit us to detect these low surface brightness features for the first time in such low mass galaxies.

The position and total brightness of each of the galaxies in our sample are listed in Table \ref{shtab}. Morphologically, they are typical of early-type dwarf galaxies, dEs. They are considerably fainter than any Virgo cluster dE sample galaxies studied by \cite{Lisker06}. VCC~1447 is the least bright galaxy in our sample. It has total brightness of M$_{g}$ =  $-$11.71~mag\footnote{In this work we assume the distance to the Virgo cluster is 16.5 Mpc, and the corresponding distance modulus and scale factor are 31.09 and 80 pc/" respectively \citep{Mei07,Blakeslee09}.} (corresponding to a B-band magnitude of M$_{B}$ = $-$11.36 mag). For comparison, the Fornax dwarf galaxy has an absolute magnitude of M$_{B}$ = $-$12.6 mag \citep{Mateo98}. All the galaxies that we study here are located at 1.6 deg or more (corresponding to a sky-projected physical radius of 460 kpc) away from M87 (i.e. the cluster center). In Table \ref{shtab}, we provide each galaxy's global properties, and list individual distances from M87, and nearest neighbour giant (M$_{B}$ $<$ $-$18 mag) galaxies. Two of the dEs studied here, VCC~1361 and VCC~1447, have no radial velocity measurement in NED. However, as we will later discuss, they are very probable members of the Virgo cluster, based on their observed morphological, structural properties and surface brightness fluctuations distance measurements. 

\begin{table*}
 \caption{}
 \label{shtab}
 \begin{tabular}{ccccccccccc}
\hline
 Galaxy & Ra & Dec  &  m$_{B}$ & v$_{r}$ & Dist. from M87 & Dist. from M49 & Neighbour & m$_{B}$ & v$_{r}$ & d \\
        & deg & deg & mag      & km/s    & deg & deg &        & mag     & km/s    & kpc  \\
\hline
VCC~1361 & 187.8629 & 09.7334 & 17.32 & ---- & 2.66 & 1.78 & NGC~4483 & 13.17 & 0875 & 213  \\
VCC~1447 & 188.1611 & 10.7595 & 19.73 & ---- & 1.69 & 2.85 & NGC~4503 & 12.12 & 1342 & 124  \\
VCC~1668 & 189.1254 & 13.5449 & 17.74 & 1414 & 1.83 & 5.79 & NGC~4569 & 10.25 & -241 & 109  \\
\hline
\end{tabular}
\\
\\
Global properties of the dwarf galaxies in which we identify low-surface brightness shell features. The first column is the name, according to the VCC catalogue. The galaxy's RA and Dec are presented in the second and third columns, respectively. The fourth column is the B-band magnitude, which is converted from g-band magnitude, using Lupton (2005)\footnote{https://www.sdss3.org/dr10/algorithms/sdssUBVRITransform.php}  equations. The fifth column is the line of sight radial velocity obtained from NED. The sixth and seventh column show the angular distance from M87 and M49, respectively. We list the name of the nearest massive (m$_{B}$ $<$ 13.09 mag) neighbour of each galaxy and their B-band magnitude, obtained from the VCC catalogue, and their line of sight radial velocity in the eighth, ninth and tenth column, respectively. The sky-projected physical distance between our sample dwarf galaxies and the nearest giant neighbour are presented in last column.
   \end{table*}

\subsection{Numerical simulations}
We conduct numerical simulations of dwarf-dwarf mergers in order to gain deeper insight into the formation mechanism of the observed shell systems, and to see if the key features of these systems can give us insight into the nature of the interaction that formed them. Although our simulations are idealised, they are well suited to achieving our goal -- to qualitatively improve our understanding of the merger process that gives rise to such features, rather than to quantitatively reproduce their exact observed properties.

We perform a series of numerical simulations of mergers between two early-type dwarf galaxies. Each dwarf is modelled as a spherical dark matter halo with an NFW profile (\citealp{Navarro96}). Our standard model dwarf galaxy has a virial mass of 2$\times$$10^9$~M$_\odot$, a concentration parameter of $c$=10, and a virial radius of 20~kpc. The dark matter halo consists of 5 million equal mass dark matter particles. Within each dark matter halo we include a spherical stellar distribution of stars following a Hernquist density profile (\citealp{Hernquist90}). The choice of a dispersion dominated distribution of stars is not unreasonable, given the typically thick shapes of early-type dwarf galaxies of this luminosity (\citealp{Ruben2016}). Our standard model has a stellar mass of 1$\times$10$^7$~M$_\odot$, an effective radius of 0.38~kpc, and is radially truncated at 1~kpc. The effective radius of the model galaxy is entirely consistent with galaxies of this luminosity (\citealp{Ferrarese06}; \citealp{Janz2009}; \citealp{McConnachie12}). Our standard model is roughly twice the mass of VCC~1447, and a few times lower in mass than VCC~1361 and VCC~1668. However as these are purely N-body simulations, we note they are scaleable to match all three observed galaxies. For example, if the mass were scaled up by a factor of 10, the simulations would produce the same results, only occurring $\sqrt{10}$=3.16 times more rapidly. However, more massive
galaxies are expected to be larger. Therefore, if we scale up the mass by a factor 10, while simultaneously scaling up the size by a factor of $\sqrt[3]{10}$=2.15, then the simulation results will be identical, with no scaling of time required. We confirm that our standard model dwarf galaxy sits on the observed scaling relations of early-type galaxies  \citep[e.g. the fundamental-plane as shown in the Fig 17 of][]{Toloba14}, and remains on the observed scaling relations even if scaled up by as much as a factor of $\sim$100 in mass.

The stellar component consists of 250 thousand equal mass stellar particles. This large number of star particles enables us to detect faint shells as more particles will fall within each shell. We conduct merger simulations with a range of mass ratio. In our 1:1 mass ratio merger simulation, the standard model merges with an identical counterpart model. However, we also conduct mergers where the standard model merges with a lower mass galaxy (total mass of 75$\%$, 66$\%$, and 50$\%$ the mass of the standard model). Thus we test mergers with a mass ratio of 1:1, 1:1.3, 1:1.5, and 1:2 (progressively more minor mergers). For model galaxies with lower masses than the standard model, we ensure the stellar to dark matter halo mass ratio remains constant at 0.5$\%$, and reduce the number of particles so as the particle masses of the two interacting galaxies remain equal. As lower mass galaxies tend to be smaller, we reduce the size of the stellar distribution as we reduce the total mass. This is achieved by reducing the effective radius and cut-off radius of the stellar component by a factor of a cube root of the fraction the mass has been reduced from the standard model. 

Initially each galaxy is placed 10~kpc apart. Then, each galaxy is given a velocity directly towards the other galaxy. We choose to model only head-on collisions (with an impact parameter of zero), as off-center collisions do not produce the aligned, interleaved shells that we observe in our sample \citep{Quinn84}. The escape velocity of the combined dwarf galaxies is approximately 50~km/s. Therefore, as we wish to model the merger scenario, we choose an initial relative velocity of 10 km/s, which is sufficiently lower than the escape velocity to ensure a merger will occur (N.b. We also consider a fly-by scenario, with a 100 km/s initial relative velocity in Section \ref{weakinteractsect}). Although our choice of initial separation and initial relative velocity, at first glance, seems very restricted, in practice, we find that these arbitrary choices have very little influence on our main results. Increasing the initial separation simply causes the two galaxies to collide at higher velocity, and so is equivalent to increasing the initial relative velocity. Meanwhile, in practice we find that setting a higher initial relative velocity merely prolongs the time we must wait for dynamical friction to cause the dwarfs to coalesce, without changing the properties of the shells that are produced by the merger (see following section for more details). 

The publicly available initial conditions set-up code, {\sc{dice}} (\citealp{Perret14}) is used to build all galaxy models. All models are initially evolved in isolation, and found to have excellent long term stability. Numerical simulations are conducted with the adaptive mesh refinement code {\sc{ramses}} (\citealp{Teyssier02}). By using an adaptive mesh refinement code, we can model the galaxy merger with very high star particle numbers, without significantly increasing simulation run-times. The high star particle numbers enable us to better detect low surface brightness features. The total simulation volume has a box size of 60~kpc which is sufficient to contain the stellar and dark matter particles of the merging galaxies over the duration of the simulation. We ensure the center of mass of the two galaxies remains at the center of the box throughout the duration of the simulation, so as to avoid unwanted numerical edge-effects from the border of the box. The minimum cell size reached in the simulations is 59~pc which is sufficient to well resolve the galaxies down to their centers. We evolve all models for 2.25~Gyr, which is easily enough for the galaxies to fully coalesce, form shells, and to continue to do so until their shells are no longer perceptible. The merger of the two dwarf galaxies is modeled in isolation, as the observed interleaving of the shells is focussed about each galaxy's center. This demonstrates that the gravitational potential of the galaxy, itself, dominates the dynamics of the stars within the shells. The results of these simulations will be discussed alongside the observations, in order to better interpret the observed shell systems and their host galaxies.

\section{Observed and simulated shell properties}

\subsection{Shell distribution}
In Figure \ref{vshel} we show a true color image, with an arcsin contrast. The true color image is prepared from the NGVS $g-, i-, z-$band images. The choice of an arcsin contrast is with the aim to obtain the best possible view of both low surface brightness shell features, and the main body of the galaxy. The shell features are clearly visible in all cases, without the need for galaxy model subtraction, or the application of unsharp masking. 

\begin{figure}
\includegraphics[width=7cm]{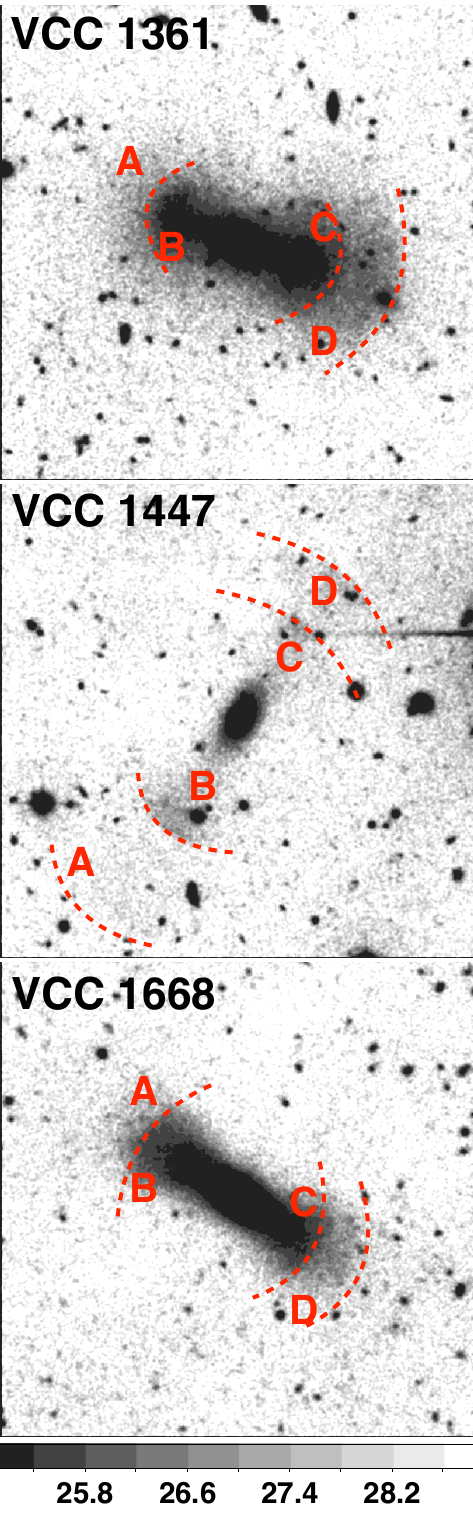}
\caption{The $g-$band surface brightness map. The positions of shell regions are located by eye where the curves, drawn by eye, highlight the positions of the shells, and correspond to breaks in surface brightness. The field of view is similar to Figure \ref{vshel}. We note that label A in VCC1361 and VCC1668 does not denote a shell-like break in the profile, but indicates a region where a low surface brightness stellar light is visible.}
\label{sfbmap}
\end{figure}

VCC~1447 contains the most extended shell system in our sample. Shells A, B, C and D are observed at distances of 17.3, 10.4, 8.8 and 14.6 times the galaxy's geometric effective radius, respectively, as measured from the galaxy's center (see Figure \ref{sfbmap} for identification). The shell arcs of VCC~1361 and VCC~1668 are closer to the center of their galaxies, and higher surface brightness, than in VCC 1447. Interestingly, we find that the shells are positioned at similar radii in VCC~1361 and VCC~1668 -- they are located at approximately 1 and 4 of Re distance from the center of their galaxy. The shells of VCC~1361 are less symmetric, and more fluffy, compared to others galaxies. Shell `A' is the broadest and faintest in VCC~1447, with a median surface brightness of $<$$\mu_{g}$$>$ = 28.48~mag~arcsec$^{-2}$, which is near the detection limit of the NGVS. It is not detected in the $i-$band image. We show the NGVS $g-$band surface brightness map in Figure \ref{sfbmap}. All shell features are fainter than $\mu_{g}$ = 25~mag~arcsec$^{-2}$.

We highlight the main shared features of the galaxies and their shells systems in the following:
\begin{itemize}
\item We see the presence of between 3 and 4 shells in each galaxy. The shells tend to have well defined edges, and there are large separations between concentric shells.
\item The shell systems tend to be fairly symmetrical about the galaxy center in position and brightness. The brightest, higher surface brightness, shells tend to be located closer to the center of the galaxy. 
\item The stellar body of the galaxies tends to be elongated and elliptical.
\item The shell systems are closely aligned with the semi-major axis of the elongated body of the galaxy. According to classification scheme of shell systems proposed by \cite{Prieur90}, all three galaxies in our sample can be classified as type-I systems.
\end{itemize}
We now consider the significance of these key features for understanding their formation process by comparison with our numerical simulations.

\begin{figure*}
    \centering
    \includegraphics[width=\textwidth]{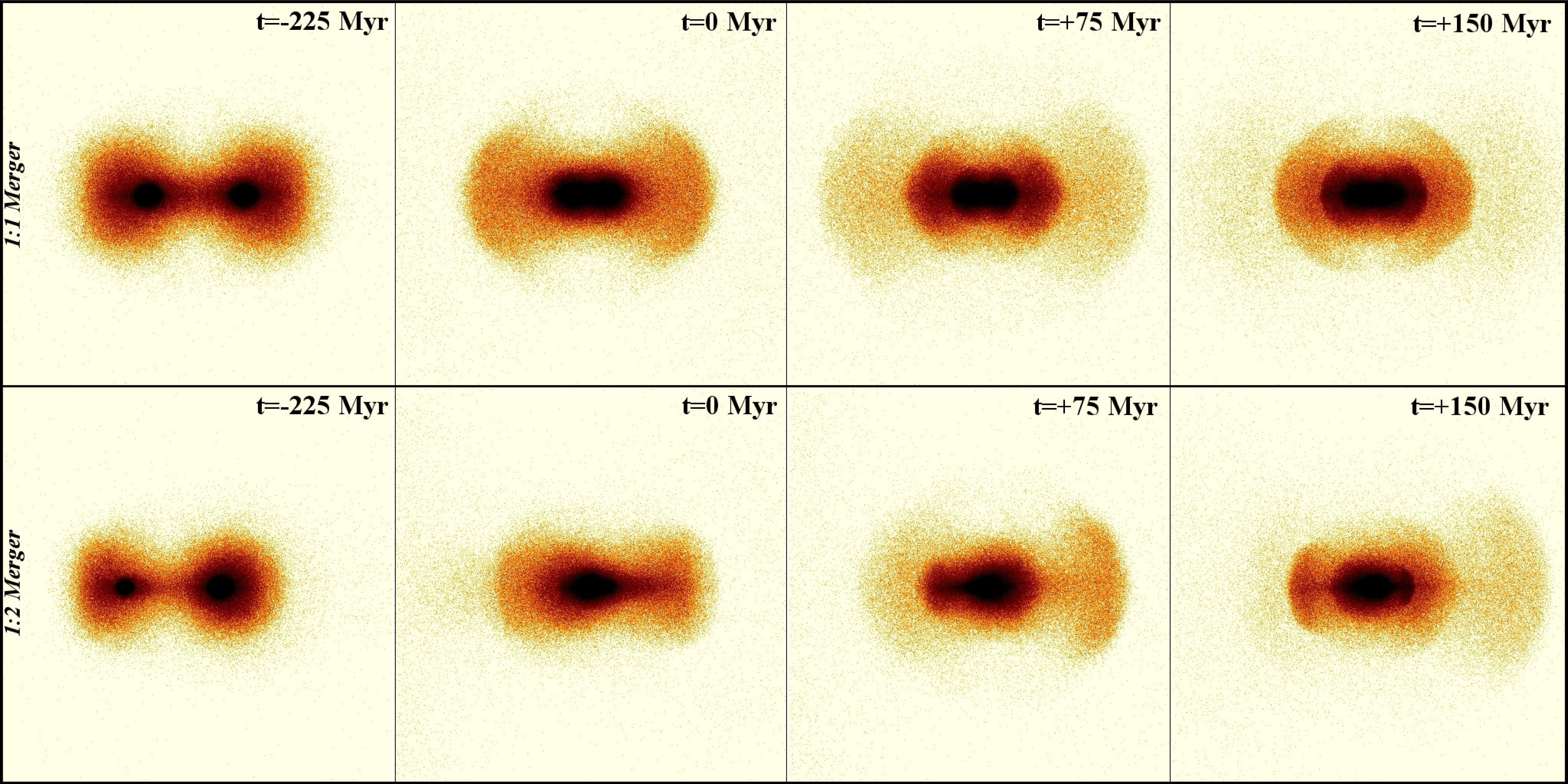}
    \caption{Time sequence (evolving from left to right) of the stellar particle distribution from prior to the merger (t=-225~Myr), until several successive shells have formed (t=+150~Myr). The top row is the 1:1 mass ratio merger, the bottom row is the 1:2 mass ratio merger. The time of the snapshot is measured with respect to the moment when the two galaxies coalesce. The box size of each snapshot is 9~kpc. Following coalescence, the galaxy is shown at 75~Myr intervals, which is the time period on which each new set of shells appears.}
    \label{snapshotsfigure}
\end{figure*}

In Fig. \ref{snapshotsfigure}, we consider the evolution of the stellar component of our model galaxies as they evolve through the merger process, during the numerical simulations. The 1:1 mass ratio merger is shown in the top row, and the 1:2 mass ratio merger is shown in bottom row. The two galaxies approach each other along the x-axis of the plots. The time of the snapshots evolves from the left to the right, as shown in the upper-right panel of each snapshot, and is calculated with respect to the time at which the two galaxies fully coalesce. Although not explicitly shown in the figure, we note that, before coalescence (t=0~Myr), the two model galaxies collide and pass through each other four times. With each close passage their subsequent peak separation is reduced by dynamical friction. However we see very few stripped stars (and thus no shells form) until final coalescence occurs. The fact that our model galaxies are dynamically hot, dispersion-supported systems makes them fairly robust to producing stellar streams during the pre-coalescence phase (\citealp{Feldmann08}). Thus these initial galaxy-galaxy collisions are of a little consequence for shell production, and simply set the scene for the final coalescence. As a result, we find that the initial relative velocity of the galaxies is not a significant parameter controlling shell production. A higher relative velocity simply causes a greater number of passages, and thus just delays the instant of coalescence.

The first set of shells appears when the galaxies coalesce (see Fig. \ref{snapshotsfigure}, `t=0~Myr' panels). Initially, a new set of shells appears every 75~Myr, and so we show each snapshot, following the merger, at $\sim$75~Myr intervals (e.g. one at t=0~Myr, two at t=+75~Myr, three at t=+150~Myr). We note that this process continues beyond the latest snapshot shown in Fig. \ref{snapshotsfigure} and, by roughly t=+500~Myr, there are so many shells that distinguishing them becomes very difficult, and they blend into one another. Furthermore, the total number of stripped stars is constant, and with each new shell, the number of stars in any individual shell is reduced, diluting the steps in luminosity between each shell. In Fig. \ref{1Gyrfigure}, we show the 1:1 mass ratio merger model at t=+1000~Myr (i.e. one Gyr since coalescence). By this time, the large numbers of shells have become so blended that the radial luminosity profile becomes smooth, with no sharp breaks at all. {\it{The limited time for which clearly defined, and well separated shells, can be seen in the simulations suggests that the observed galaxies must have suffered a recent coalescence, within the last few hundred Myrs.}} 

\begin{figure}
    \centering
    \includegraphics[width=8cm]{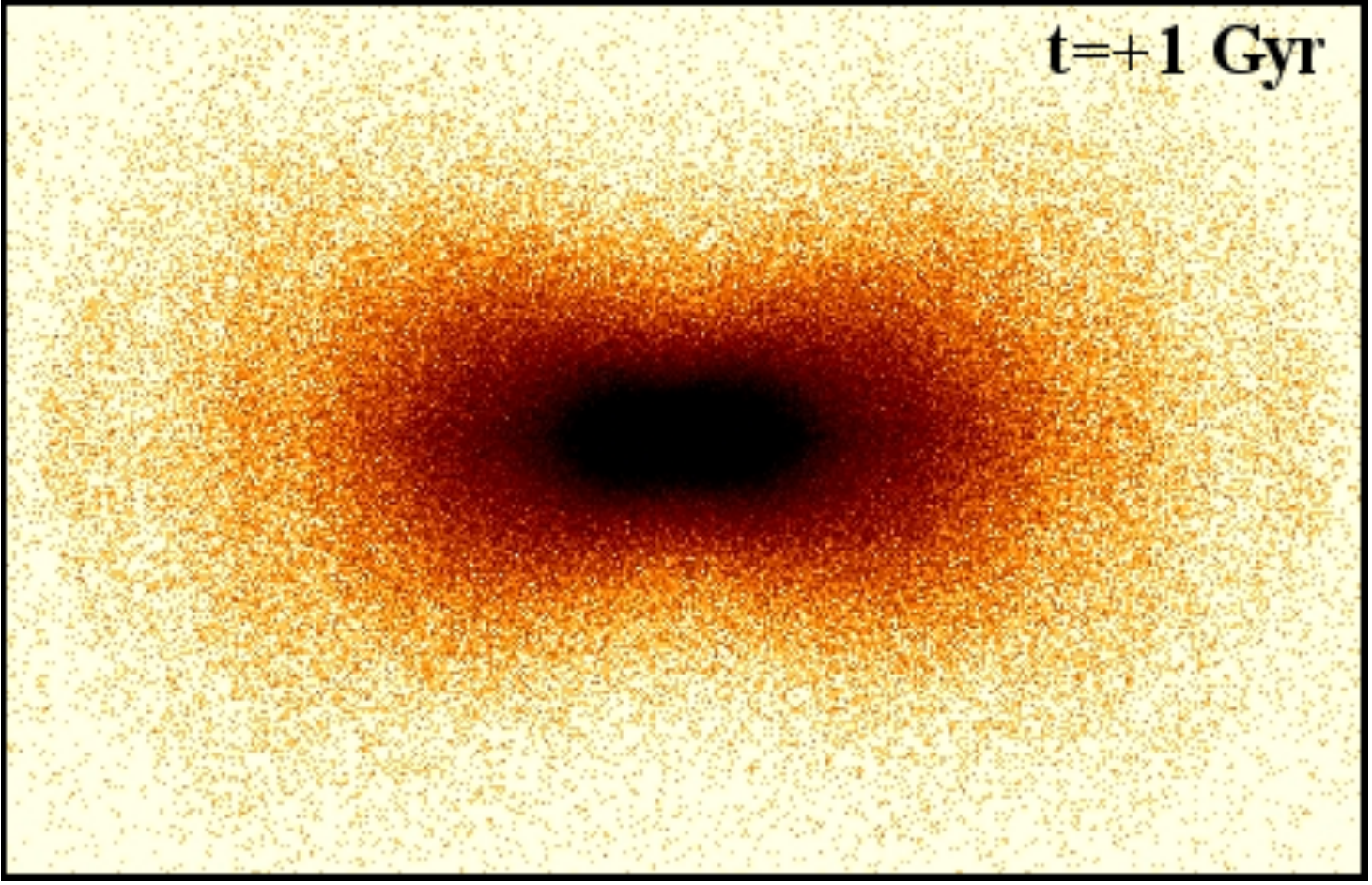}
    \caption{The merger remnant shown 1 Gyr after galaxy coalescence. Box size is 5 by 9 kpc.}
\label{1Gyrfigure}
\end{figure}

In fact, the reason that clear, well-defined shell features can only be seen for short time-scales in our dwarf models, is intrinsically linked to the fact that the pre-collision dwarf models are supported by a significant component of dispersion. In \cite{Quinn84} it is stated that, with increasing number of shells, the widths of individual shells become comparable to the shell separations, and at this point the shells ``wash-out, and cease to be sharp features". Because our dwarf models are dispersion supported systems, the shell widths are thicker (see Equation 19 of \citealp{Quinn84})\footnote{Also compare Figure 13 and 14 of \cite{Dupraz86}, where the effect of a high velocity dispersion on shell thickness can be clearly seen. It is also notable that the surface brightness profile of the shells in these figures, produced by the merger of an elliptical instead of a spiral, is a better qualitative match to our observed shells.}, causing the wash-out to be more severe. From their Fig. 12, we estimate that wash-out of the shells, as described in their paper, would occur when there is no more than $\sim$6 shells, which is less than half a Gyr after the merger, assuming 75~Myr per set of shells. More recently, \cite{Feldmann08} studied the duration for which merger features are visible when an equal mass merger occurs between two elliptical galaxies, and found that intermediate or strong tidal features were visible for less than 100~Myr\footnote{\cite{Feldmann08} did not specifically focus on shell features, but their definition of `tidal features' should have also included the light from stars in shell features.}. Meanwhile in a 1:4 or 1:10 mass ratio merger between an elliptical and a spiral, intermediate and strong tidal features remained visible for $\sim$1.5~Gyr. Therefore, our derived timescale of a few hundred Myrs since coalescence is not inconsistent with previous studies in the literature, once the dispersion support of the pre-collision galaxies is taken into account. We also note that, at these luminosities, most early-type galaxies are not highly flattened, and so are expected to be significantly supported by dispersion. As a result, similarly short timescales for detecting merger features in dwarf-dwarf mergers may be very typical for galaxies of this luminosity.

Comparing the top row (1:1 mass ratio merger) to the bottom row (1:2 mass ratio merger) of Fig. \ref{snapshotsfigure}, it is clear that by reducing the mass of one of the galaxies by only a factor of two, a strong asymmetry (about the y-axis) is introduced into the family of shells. In fact, the position of the shells is only mildly affected by the mass ratio, but the number of stars found in each shell changes significantly, causing the asymmetric appearance. Although not explicitly shown here, we note that in the 1:1.5 mass ratio merger the asymmetry is still visible, but in the 1:1.3 mass ratio merger (which is nearly 1:1 mass ratio) the shells begin to look very close to symmetrical once more. As the observed galaxies present shell structures that appear fairly symmetrical, this could suggest that a mass ratio merger ratio of between 1:2 and 1:1 is required, and we refer to such mergers as near equal mass mergers herein. We note that VCC 1361 perhaps qualitatively resembles the 1:2 merger mass ratio model more than the equal mass merger model.

Once the stars have been released, it is their orbit within surrounding potential well which dictates the shape of the shells. The concentric shape of the observed shells about the galaxy core thus indicates that the gravitational potential well in which the stars orbit has been dominated by the coalesced dwarf galaxy's own self-gravity since the galaxies fully merged. This rules out the strong influence of the cluster potential, or a high speed tidal encounter with another cluster member, at least in the last few hundred  Myr.\footnote{Located at beyond 450~kpc or nearly halfway out of cluster virial radius, these galaxies may not feel strong tidal interaction from the central potential see \cite[][Figure 10]{Smith15}}

The requirement for a near equal mass ratio merger is further supported by the elongated shape of the cores of the observed galaxies. For example, the bright core of VCC~1447 has an ellipticity of 0.5\footnote{We define the core as a regular morphological part of galaxies, see Figure \ref{sfbmap}}. We similarly measure the final ellipticity of the stellar cores of the modelled merger remnants and find a clear trend for increasing eccentricity with mass ratio of the merger. This can be physically understood -- our model galaxies are initially spherical. Therefore if the merger is too minor, the final coalesced system will remain near spherical. In fact, the merger mass ratio of 1:2, 1:1.5, 1:1.3 and 1:1 (progressively more major) has an ellipticity of 0.41, 0.43, 0.44, and 0.52 respectively (progressively more elongated). Actually, only the 1:1 mass ratio succeeds in reaching the observed ellipticity. However it could also be argued that the pre-merger galaxies may have been elongated before the merger. But if they were elongated, it would be necessary for their position angle to be closely aligned with their orbital plane, otherwise the observed shells would not appear so well aligned with the position angle of the central core (as observed in all three dwarf galaxies). Furthermore, without a close alignment of their position angle with the orbital plane, the merger would also cause an initially elongated stellar distribution to thicken. The problem with this scenario is that there is a low probability of alignment between the position angle and orbital plane, given the wide possible range of orbital configurations possible. Therefore elongation of the core as a result of a near equal mass ratio merger is much more probable, and simultaneously explains the shell symmetry, core ellipticity, and alignment of the shells with the position angle of the core, without requiring any special and/or unlikely configurations.

\subsection{Surface photometry}\label{surface}

Surface photometry of very low surface brightness objects is notoriously challenging. We use a similar procedure to clean the NGVS images as in our previous study \citep{Paudel13}, where we studied similarly faint structures. A ring median filter is applied to each image, to remove foreground stars and compact background galaxies \citep{Secker95}, and the residuals of extended objects were manually masked with the IRAF task $imedit$. The errors in the photometric measurements are, to a large extent, determined by those in the sky determination. Although a master sky background is subtracted from each image, for our aperture photometry measurements we also sample the sky background count, using a similar procedure as applied in \cite{Paudel13}.

After preparing the cleaned, and sky-background subtracted image, we used the IRAF $ellipse$ task to fit galaxy's isophotes to ellipses. To avoid uncertainty in ellipse fitting, due to foreground and background objects, we prepare masked images manually. The center and position angle were held fixed, and ellipticity was allowed to vary during the ellipse fit. We calculate the centers of the galaxies using the IRAF task $imcntr$, and the position angles are determined using several iterative runs of $ellipse$ before the final run.

Due to rather the irregular morphology it is difficult to approximate the observed light distribution with a single parametric function, and then use this to derive the global photometric parameters. To circumvent these difficulties, we instead select the magnitudes and half light radii derived from curve of growth photometry in the NGVS images. A detailed description of the photometry of the NGVS images will be presented in Ferrarese, et al. 2016, in press. We present the $u$, $g$, $i$, and $z$ band magnitudes and geometric half light radii of our sample galaxies in Table \ref{phtab}.  We convert the $g-$band magnitude to  stellar mass, using the mass-to-light ratios obtained from \cite{Bell03}, based on $g-i$ colors.

We now attempt to calculate the fraction of the galaxy's total light contained in shell feature (L$_{\rm{shell}}$/L$_{\rm{tot}}$). We subtract the best fit ellipse model and sum up the residual flux (L$_{\rm{shel}}$) which is predominantly light associated with shell features. The value of L$_{\rm{shell}}$/L$_{\rm{tot}}$  is 0.13, 0.35 and 0.11 for VCC 1361, VCC 1447 and VCC 1668, respectively. In all cases, the amount of light left over after subtraction of the model galaxy is a significant fraction of the total. In fact, the clear visibility of the shell features, even without subtracting a model for the galaxy, emphasizes that this fraction is relatively very high. This may provide secondary evidence that a major merger must have occurred.
For example, if a relatively minor 1:20 mass ratio merger occurred, and all of the less massive galaxy's stars are distributed in the shell only after the merger (an extreme case, as not all of the stars would be left in side shells), then L$_{\rm{shell}}$/L$_{\rm{tot}}$ could not exceed 0.05, where as all three dwarfs exceed L$_{\rm{shell}}$/L$_{\rm{tot}}$=0.1. Following this reasoning, VCC 1361 or VCC 1668, with L$_{\rm{shell}}$/L$_{\rm{tot}}$$>$0.1, must have suffered a more major merger than a 1:10 mass ratio. Similarly, the merger ratio of VCC 1447  must have been more major than 1:3. In fact, in a near equal mass merger, both galaxies involved may contribute stripped stars, enabling even higher resulting values of L$_{\rm{shell}}$/L$_{\rm{tot}}$.

\begin{table*}
\centering
 \caption{}
 \label{phtab}
 \begin{tabular}{ccccccccc}
\hline
Galaxy & m$_{u}$ & m$_{g }$ & m$_{i}$ & m$_{z}$ & m$_{NUV}$ &  M$_{*}$& Re    &  $\epsilon$ \\
        & mag & mag & mag & mag & mag &  $10^{7} \times$M$_{\sun}$ & kpc &    \\
\hline
VCC1361 & 18.13 & 16.97 & 16.41 & 16.07 & 21.00 &  6.9 & 1.62  &  0.43 \\
VCC1447 & 20.54 & 19.38 & 18.57 & 18.52 & --    &  1.3 & 0.27  &  0.36 \\
VCC1668 & 18.64 & 17.38 & 16.75 & 16.74 & 21.68 &  5.5 & 0.80  &  0.48 \\
\hline
\end{tabular}
\\
Results of surface photometry in NGVS images. In columns 2-5, we list the magnitude from the NGVS source catalogue, which are derived from curve of growth photometry. The sixth column is the NUV magnitude from \cite{Voyer14}. We obtained stellar masses from the $g-$band magnitude, assuming a mass-to-light ratio derived from the $g-i$ color, using equations in \cite{Bell03}. In columns eight and nine we list geometric half light radius and average ellipticity of the galaxies, respectively.
\end{table*}

\subsection{Colors}
In Figure \ref{sfbmap}, we schematize the location of the shell features, and label individual features. The red dotted curve show the shell outlines in order to highlight where the break in surface brightness occurs. The approximate radial distances from the galaxy center to each arcs is listed in column 3 of Table \ref{feature}, in units of each galaxy's geometric effective radius. The exact location of the edge of shell A in VCC 1361, and VCC 1668, is not as clear as in other cases, so we do not list its position in Table \ref{feature}.

When features are extremely low surface brightness (i.e. 100 times fainter than the sky brightness), photometry is not trivial. As a result, we first mask all unrelated foreground and background objects, and we smooth the images using a gaussian kernel, with a 3-pixel sigma. We then select each region, as designated in Figure \ref{sfbmap}, and measure the mean surface brightness, and $g-i$ color index. The $g-i$ color index is measured in two different ways; one from the integrated flux in the region (i.e. a mean color), and another from the median surface brightness. We find that the results are consistent with each other.  The $g-i$ color index derived from the median surface brightness is listed in the last column of Table \ref{feature}. All three galaxies, and their shells, show red optical colors (e.g. $g-i$ = 0.6 - 0.9 mag), see Table \ref{phtab} \& \ref{feature}. According to our simulations, a small number of well defined, clearly separated shells (as observed) is strong evidence for a recent merger, occurring in the last few hundred Myrs since the system coalesced. The requirement for a recent merger, but where the coalesced system displays red optical colors, is highly suggestive of a dry merger. Neither galaxy could contain gas, otherwise the merger would likely have induced a starburst that would still appear blue so shortly after coalescence. There is also no indication of a radial color gradient, or the presence of dust lanes, which are common indicators of a wet merger.

\begin{table}
\caption{}
 \label{feature}
\begin{tabular}{|cccccc|}
\hline
Galaxy  & Shell &  Pos  & $<$$\mu_{g}$$>$ & $<$$\mu_{g}$$>$ err  & g-i\\
\hline
  & & R/Re(kpc) & mag/arcsec$^{2}$ & mag & mag  \\
\hline
VCC1361 & A  & ---  &   26.8 &  0.1   &  ---    \\
VCC1361 & B & 1.0(1.6)  &   25.5 &  0.1   &  0.7    \\  
VCC1361 & C & 1.4(2.2) &   25.4 &  0.1   &  0.7    \\  
VCC1361 & D & 2.1(3.4) &   26.2 &  0.1   &  0.8    \\  
VCC1447 & A & 17.3(4.6) &   28.6 &  0.2   &  ---     \\         
VCC1447 & B & 10.4(2.8)  &   27.1 &  0.1   &  0.8    \\
VCC1447 & C & 8.8(2.3) &   27.1 &  0.1   &  0.7    \\
VCC1447 & D & 14.6(3.9) &   27.3 &  0.1   &  0.8    \\
VCC1668 & A & ---  &   27.9 &  0.1   &  0.7    \\
VCC1668 & B & 2.9(2.3)  &   25.8 &  0.1   &  0.8    \\
VCC1668 & C & 2.3(1.8)  &   25.7 &  0.1   &  0.9    \\
VCC1668 & D & 3.4(2.7) &   26.9 &  0.1   &  0.9    \\
\hline
\end{tabular}
 \\
 \\
 Properties of the shell features, as labelled in Figure \ref{sfbmap}. The third column represent the distance as function of geometric half-light radius at which the sharp break \tclr{in surface brightness is observed. Absolute values are given in parentheses.} The location of shell edges is highlighted by red dash lines, that are positioned by eye. The average surface brightness, given in the 3rd column, IS derived from the median pixel flux in each region. In the fourth column, we present the $g-i$ color of each feature.
   \end{table}

\begin{figure}
    \centering
    \includegraphics[width=8cm]{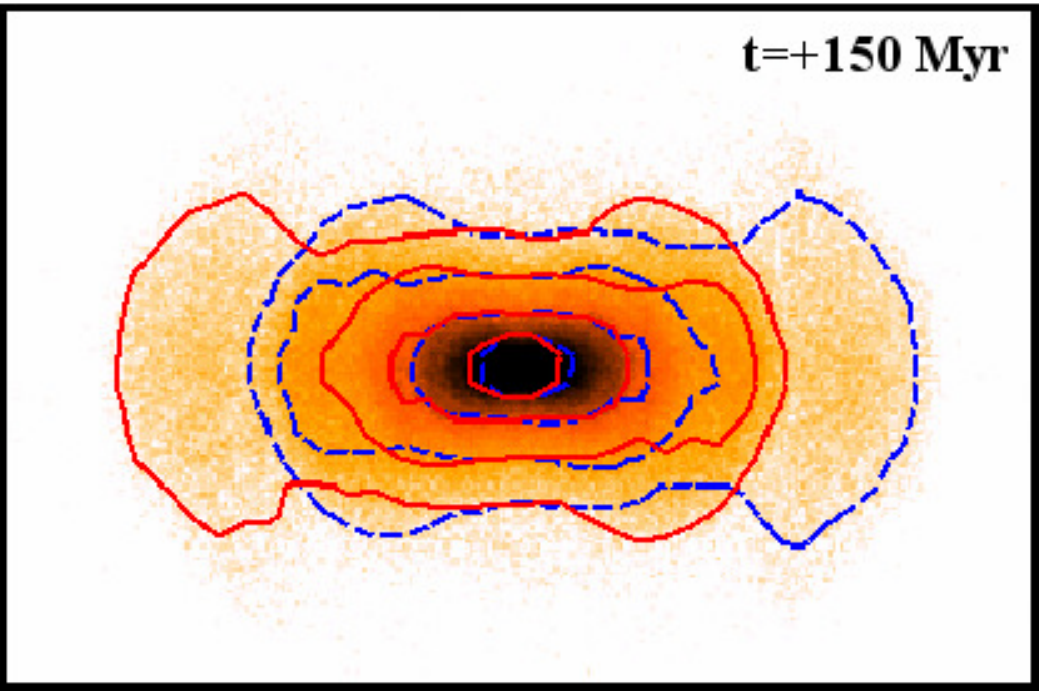}
    \caption{Stars of the 1:1 mass ratio merger shown at t=+150~Myr since coalescence (orange points, darker shade means higher mass density). Surface mass density contours are arbitrarily chosen to highlight the shell edges. Red solid contours indicate shells produced by stars from one galaxy, and blue dashed contours indicate shells produced by stars from the other galaxy. The box size is 5~kpc by 3.25~kpc.}
\label{shellfigure}
\end{figure}

While conducting the simulations, we additionally noted that each shell could originate from one or the other of the two progenitor galaxies. In Fig. \ref{shellfigure}, we show the stellar distribution of the 1:1 mass ratio merger 150~Myr after the final coalescence. On top of the stellar distribution, we overlay surface density contours. We arbitrarily choose contour values in order to best highlight the positions of shell edges. Solid red contours trace out the stars originating from one galaxy, meanwhile dashed blue contours trace out stars originating from the other galaxy. This exercise reveals that, in some cases, a shell generated by the first galaxy is often located directly beside a shell generated by the other galaxy. In fact, successive shells may alternate as to which galaxy they were initialized in. Furthermore, although the total stellar distribution is symmetric about the y-axis, the contours of stars from an individual galaxy are far from symmetrical. However, the blue contours are the mirror image of the red contours, and there are equal masses of stars in the red and blue contours, hence the combined shells are symmetric. This is a natural result of a near 1:1 mass ratio merger -- both galaxies release near-equal quantities of stars simultaneously, and as a result both galaxies contribute equally to forming the final shell population, which is symmetric once the shells of both galaxies are combined.

As a result, potentially we might be able to detect differences in the stellar populations of different shells, depending on which galaxy they have originated from. Therefore, in the following, we attempt to detect differences in the colors of shells.

We find no significant difference in color between the shells, and their galaxy's main stellar body (see Table \ref{phtab} \& \ref{feature}), at least as far as the signal to noise ratio permits us. However this probably suggests that the two early type dwarf galaxies involved in the merger had similar colors. Indeed, the clear red sequence seen in early-type dwarf (e.g. \citealp{Janz2009}) indicates that similar mass dwarfs galaxies are expected to have similar colors. However, as a color difference was not seen in deep imaging, then detecting a possible difference between the stellar populations of successive shells likely becomes very technically challenging, as spectroscopy of such faint structures would be highly demanding. 

\section{Discussion}

\subsection{A dry merger origin of the shells}

We further explore the origin of these low surface brightness features with the help of idealised numerical simulations. In the model we are able to reproduce the visual appearance of the shells, and galaxy, following a near-equal mass merger. The simulations provide support for a formation scenario in which a recent merger, between two near-equal mass, gas-free dwarf galaxies, forms the observed shell systems. This picture is supported by multiple lines of evidence. In the models, every 75~Myr since coalescence, a new set of shells appears. Therefore the small number of well defined shells, that are observed, indicates a recent coalescence in the last few crossing-times (in the last few hundred Myr). 

Observationally, the shells are fairly symmetric about the core of the galaxies, the core is elongated, and the shells are well aligned with the position angle of the core. All of these properties, combined, are found to naturally arise in our models, following a major merger of two near-equal mass{\bf{, gas-free}}, dwarf galaxies. The presence of these shells likely rules out the strong influence of the cluster tides on these galaxies, at least for the time since the merger occurred, as strong external tides could disrupt the shells. The fact that the shells interleave about the center of each galaxy indicates that the galaxy's potential is currently the dominating potential controlling the motion of the stars within the shell.

We see no evidence that the merger was wet. There is no radial color gradient, and we see no dust lanes. We also measure the UV-optical color of the galaxies in our sample, and place them on a UV-optical color-magnitude diagram. VCC~1361 and VCC~1668 are detected in  NUV image of the GALEX Ultraviolet Virgo Cluster Survey \citep[GUViCS][]{Boselli11}, and the reported magnitudes are 21.00 and 21.68~mag, respectively. From this we derive a NUV$-i$ color larger than 4.5~mag, for galaxies with a stellar mass less than 1$\times$10$^{8}$ M$_{\sun}$. Thus they are perfectly consistent with being red sequence galaxies (e.g. see Figure 10 of \cite{Boselli14}\footnote{In \cite{Boselli14}, galaxies with a stellar mass of 10$^{8}$ M$_{\sun}$ become red sequence galaxies when they have NUV$-i$ color redder than 3.8~mag.}), and so we see no evidence for a star burst that might have occurred if the merger had been wet, and occurred recently. These galaxies are not detected in large surveys of Neutral Hydrogen (HI), such as  ALPHA-ALPHA \citep{Giovanelli05} or HIPASS \citep{Meyer04}.

\subsection{A weak interaction origin of the shells}
\label{weakinteractsect}

\begin{figure}
    \centering
    \includegraphics[width=8cm]{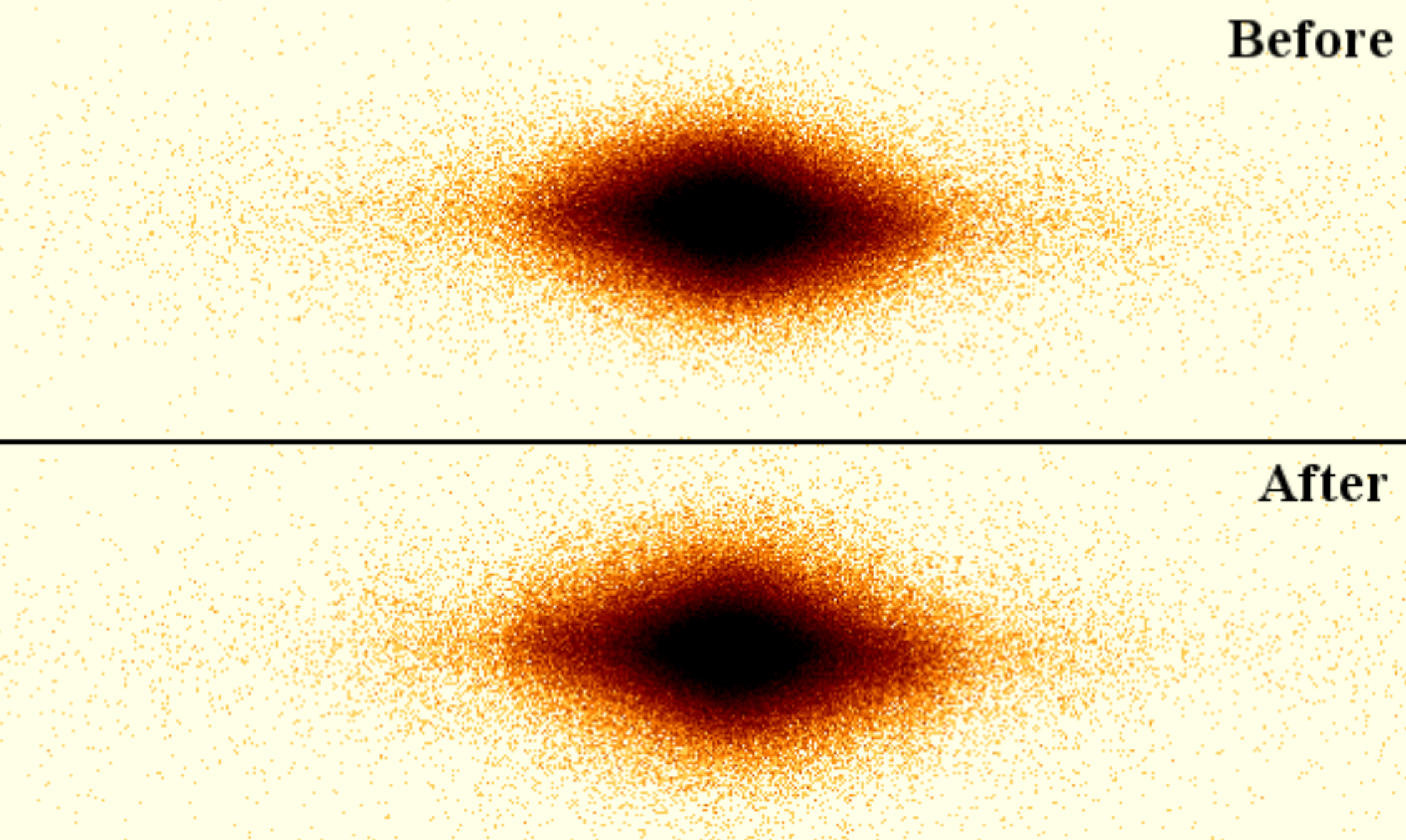}
    \caption{ Fly-by encounter with 1:1 mass ratio. We show the relaxed initial conditions (before), and about 150 Myr afterwards (after). The box size is of 3 by 9 kpc.}
\label{1Gyrfigure}
\end{figure}

We also test an alternative scenario to shell formation by mergers. In \cite{Thomson90,Thomson91}, it is suggested that elliptical galaxies may contain a thick disk component. They find that shells can be produced in the thick disk of stars, by a single weak tidal encounter with another galaxy, of 10-20$\%$ the mass of the elliptical galaxy. In contrast to the merger scenario, the interaction velocity is sufficiently high that the galaxies do not merge (i.e. a `fly-by' encounter instead). Because the thick disk is not spherical, it produces ring-like structures in the disk plane, in response to the tidal encounter. Viewed face-on to the disk, several complete rings may be generated. However, viewed edge-on, the ring structures instead appear as shell-like features, as a result of projection effects \cite[see for example Figure 4 of][]{Thomson90}.

To test this scenario, we set up a thick, exponential disk of stars, within an NFW halo. The halo properties match that of our standard model's halo. For the thick disk, we choose an exponential scale radius of 2~kpc, and an exponential scale-height of 1~kpc. We then place a second NFW halo, with 20$\%$ of the mass of the first halo, at a distance of 10~kpc, and send it directly towards the center of the first halo, with a relative velocity of 100~km/s. The escape velocity of the first halo is roughly 20~km/s. Therefore, the encounter is at sufficiently high velocity to ensure a fly-by occurs. The encounter impacts the disk edge-on, although \cite{Thomson90} find the inclination of the disk is of little consequence to the end results.

In practice, we find we are unable to produce ring shaped features, or shell-like features in our fly-by models. In \cite{Thomson90}, a weak tidal encounter is preferred. In fact, the response of our thick disk models to a weak-interaction, fly-by encounter is almost imperceptible.  As we fail to see a clear response, we increase the mass of the fly-by halo until the two galaxies have equal mass, in order to produce a much stronger tidal encounter. Now the thick disk can be seen to respond to the encounter. During the fly-by it contracts, and then re-expands once the fly-by has occurred. The stars involved in the re-expansion form a low surface brightness envelope surrounding the disk, but we see no indication of ring-like structures, (see Figure \ref{1Gyrfigure}). Thus we find it challenging to reproduce the observed shells features using the weak interaction scenario, and find that the major merger scenario is much more successful. In addition, we note that, even if the fly-by scenario were successful at producing shells when viewed edge-on, then we might expect to see some dwarfs surrounded by complete rings, when viewed face-on.

\subsection{Comparison with Fornax dwarf galaxy}
We observe a significantly higher number of shell features in our sample galaxies compared to the shell detection in the Fornax dwarf galaxy \citep{Coleman04}. We note that, recently, \cite{Bate15} showed that the outer shell structure, at 1.4 degree from the center of Fornax may, in fact, be a mis-identified over-density of background galaxies. {\it{Therefore there is some uncertainty about the very existence of a shell in Fornax. In comparison, there is no doubt that the shell features in our galaxies are real, and much more prominent than the Fornax,}} \citep[see][Figure 1]{Coleman04}.  We find that shells in our galaxies are distributed along the major axis of each galaxy, while in the Fornax the shell is located along the minor axis. As shown in our idealised numerical simulations, the shell alignment along the major axis provides supporting evidence for a major merger origin in our sample. Therefore inversely they could support a minor merger origin the Fornax. The shells system we observe in our galaxies can be classified as type-I. For example, visually VCC~1447 resembles a miniature version of the shell system in NGC~7600 \citep[see][Figure 3]{Cooper11}.

We find that, within the errors, there is no significant difference in color between the shell features and the galaxy main body for our galaxy sample, suggesting similar stellar populations. In comparison, the Fornax dwarf galaxy shell features are dominated by a stellar population with young ages, while the main body hosts a mix of both young and old stellar populations. However there are suggestions that the relatively blue over-dense region, inner shell feature suggested by \cite{Coleman04} is in fact a patchy star-forming region \citep{Battaglia06, Olszewski06} or recently disrupted young star-cluster \citep{Penarrubia09}. Furthermore, a detailed stellar population study shows that Fornax possesses a complex history of star formation with a significant radial age gradient \citep{Boer12}. This suggests that while Fornax may have suffered a wet merger, the mergers in our samples were more likely dry. This is further supported by the numerical modelling, which shows that the higher number of observed shells implies a recent merger has occurred. If the merger occured recently, and yet there is no indication of recent star formation, then the requirement for dry mergers in our sample is further strengthened. Fornax is comparatively round with an ellipticity of $\epsilon$ = 0.3, whereas our dwarf galaxies have the ellipticity of $\epsilon$ $>$ 0.35. In our numerical simulations, a more major merger produced a more elliptical dwarf galaxy.

Assuming the Fornax dwarf galaxy feature is a true shell, our sample represents the only additional dwarf galaxies of stellar mass range $\lesssim$10$^{8}$ M$_{\sun}$ now known to possess shell features. However, the environment in which our shells have formed is very different from that of Fornax dwarf. Fornax is located in a group environment, $\approx$138 kpc (0.5 virial radius) away from its host galaxy, the Milky-Way \citep{Saviane00}. It has a heliocentric line of sight velocity of 53 km/s \citep{Mateo98}. Meanwhile our sample galaxies are located at a projected radius of nearly halfway out from the virial radius of the Virgo cluster. The minimum projected separation between the dwarf galaxies, and their respective nearest giant (M$_{B}$ $<$ -18 mag) neighbour, is 109 kpc, which is in the case of VCC 1668 but has relative line of sight radial velocity larger than 1500 km/s.

\subsection{Cluster membership of the dwarfs with shell features}

Of the three galaxies discussed in this paper only one, VCC~1668, has a spectroscopic redshift (v = 1414 km/s) and is a confirmed cluster member. For the other two galaxies, we must rely on structural and photometric parameters to assign membership. \cite{Ferrarese16} present a detailed discussion of the criteria used to assess cluster membership for the NGVS galaxies. Such criteria include several diagnostic plots based on structural as well as photometric parameters (including scaling relations, photometric redshifts and stellar population diagnostics), and are supplemented by a visual assessment. As discussed in \cite{Ferrarese16}, the NGVS membership criteria are validated based on a number of tests, including cross correlation with spectroscopic catalogues and blind search in control (non-Virgo) fields. Based on such criteria, all three galaxies discussed in this paper are classified as cluster members. In particular, VCC~1361 and VCC~1668 are classified as ``certain" (class 0 in the NGVS classification scheme) member, while VCC~1447 is classified as a ``likely" member (class 1).

In addition to this,  we look at the results of Surface Brightness Fluctuation (SBF) analysis of the NGVS images (Cantiello et al in preparation).  The distance modulis are 31.2, 31.3 and 31.1 mag for the VCC~1361, VCC~1447 and VCC~1668, respectively. A typical systematic error is of 0.5 mag. This result is consistent with our assumption that these three galaxies are associated with the Virgo cluster, which has a mean distance of 16.5 Mpc (corresponding to a distance modulus of 31.09 mag). Given that no dense large-scale structures are located in the background of the Virgo cluster (at least within the redshift range $<$ 4000 km/s), then it is likely that these galaxies are indeed associated with the Virgo cluster, and not with a background substructure.

\subsection{Dwarf-dwarf mergers in the Virgo cluster}

We have discovered dwarf galaxies which have likely evolved through near-equal mass mergers, occurring while the galaxies are associated with the Virgo cluster. It is not impossible that the merger might have happened outside of the cluster, and they have recently fallen into the cluster. However, the stellar population properties suggest that participating galaxies were already gas-poor and non star-forming. The virtual non-existence of such type galaxies outside the cluster or group environment \citep{Geha12} thus favors our hypothesis. Relatedly, in a large sample of observed dwarf-dwarf interactions, \cite{Stierwalt15} notice that all interacting dwarf galaxies with no/little star-formation are located near massive giant galaxies and all interacting dwarf galaxies in isolated environment are actively star-forming. Furthermore, the small numbers of observed shells, combined with our numerical simulations, suggests that the merger had to occur within the last few hundred Myrs, making it challenging for the galaxy to fall from the cluster outskirts to inside the cluster in such a short time period.

As of yet, fully cosmological simulations have not reached the mass resolution that the observations are now probing. However, based on the Local group simulation results \citep{Deason14}, major mergers between sub-halos are quite rare. \cite{Klimentowski10} predicts that $\sim$10\% of dwarf galaxies of stellar mass M$_{*}$ $\approx$ 10$^{7}$ M$_{*}$ suffer a major merger but almost all these mergers occur before 7 Gyr (i.e. prior to infall into the host galaxy). The results of \cite{Angulo09}, which are based on Millenium simulation, are also of interest to this study. While the Millenium simulation is unable to reach the resolution found in the Local Group simulations, they do include clusters of mass similar to Virgo. From their Figures 5 \& 10, we can calculate that only $>$0.2\% of the least massive subhalos have mergers more major than 1:3 since infall into their host. Furthermore, the merger rate (major {\it{and minor}}) of their least massive halos (10$^{10}$M$_{\sun}$) in Virgo cluster mass halos is still only about 0.2 mergers per subhalo, every 1 Gyr. Based on these results, we conservatively estimate that we can expect less than 1\% of dwarfs to have had a major merger since infall into Virgo, and if we require a very recent merger, the fraction might be even less.

From the observational side, we also wish to emphasize that, here, we have only selected dwarf galaxies of early-type morphology, with easily identified shell features, and we have found three. In the deep NGVS imaging, we find many other types of dwarf galaxies with shell-like features, including a number of star-forming, gas-rich, BCDs. The nature of these other galaxies will be explored in Zhang et al. (in preparation). It is not clear if all of them are the product of dwarf-dwarf mergers. However, we note that majority of the shell-like features in the other dwarf galaxies, that are not studied here, are different in term of the shell-like features morphology, and surface brightness contrast, compared to this sample. In any case, under the assumption that shell features originate from dwarf-dwarf mergers, our discoveries suggest that dwarf-dwarf mergers can occur, -in the cluster environment-, despite the fact that the velocity dispersion between galaxies maybe $\sim$1000~km/s.

The results in this paper support a scenario in which a major-merger between two low mass early-type galaxies, which are associated with the Virgo cluster, has recently occurred. However, given the uncertainties associated with the surface brightness fluctuation method, it is impossible for us to acertain the clustocentric radius of our sample of dwarf galaxies with sufficient accuracy to distinguish between if they are deep within the cluster core, or in the cluster outskirts, perhaps infalling for the first time. {\it{If they are, in fact, deep inside the cluster, then the presence of clear merger features in all three dwarfs is truly remarkable}}, considering that \cite{Behroozi2014} finds that, in $\Lambda$CDM, most major mergers end for infalling halos at greater than four virial radii away from the cluster. It is even more remarkable, if we assume that all three mergers had to occur recently, in the last few hundred Myrs. Therefore this hypothesis might favor that these galaxies are in the cluster outskirts, infalling for the first time (although see below). 

Intrinsically, low luminosity dwarfs, such as in this sample, have low masses and so low escape velocities. We calculate the escape velocity of a single, pre-collision model dwarf to be $\sim$30~km/s. Therefore if interaction velocities between two dwarfs are significantly higher than this (e.g. $\sim$50~km/s) then, instead of a merger, a galaxy-galaxy fly-by will occur (see modelling of fly-bys in Section 4.2). The high velocity dispersion of cluster galaxies ($\sim$1000~km/s) essentially rules out a chance encounter between two cluster galaxies, that were previously unassociated with each other. Instead some form of pre-association between the two pre-collision dwarfs is required. 

One possibility is that the pre-collision dwarfs were both members of a small group. However, even in small groups, collision velocities of less than $\sim$50~km/s may be very rare. If both pre-collision dwarfs were actually satellites of an intermediate mass host galaxy \citep[as in the case of NGC~4216 -- see][]{Paudel13}, then the lower halo mass of the host might be more conducive to generating lower velocity collisions between satellites. The presence of a nearby, more massive galaxy could also help explain how both pre-collision dwarf galaxies were gas-free, given that isolated galaxies of this mass are expected to be late-type. However, Table \ref{shtab} illustrates that, although all of the dwarfs have a giant companion that is close in the projected distance, we cannot confirm if the association is real, given that two of the dwarfs do not have a spectroscopic redshift, which could be used to confirm any association. The one dwarf that does have a spectroscopic redshift, VCC~1668, is clearly not at all associated with its nearby giant companion, given that the radial velocity difference is more than 1650~km/s. It is possible that the dwarfs were originally associated with a host galaxy, but have since been tidally stripped from that host by the cluster tides. However this would require a deep, plunging passage of the cluster core, which is not consistent with our previous argument that the galaxies are likely in the cluster outskirts, infalling for the first time. There is also no evidence of strong cluster tides, nor the tidal field of a host galaxy, currently acting on the dwarfs, as such a strong tidal field would likely disrupt the observed, interleaved shape of the shells about each galaxy's center. Although, in principle, the pericenter passage of the cluster core may have happened in the past, so removing the dwarfs from their host, but somehow avoiding to disrupt the bound relationship between the two dwarfs in the process. 

Alternatively, perhaps the pre-collision dwarf galaxies were born as a binary pair. In this case, their low interaction velocities are present at birth, without the need for a nearby host. Indeed a number of binary dwarf galaxies are already discussed in the literature including the LMC/SMC, and Leo-IV/V (\citealp{Blana2012}), although as Local Group members, they are in the presence of two giant host galaxies. In fact, the majority of the interacting dwarf galaxies in \cite{Stierwalt15} are located in isolated environments, without the need for the presence of a nearby giant host galaxy. Nevertheless, without the presence of a nearby giant host galaxy, it is not obvious how two low luminosity, binary dwarfs became gas poor, prior to their merger.

\subsection{Conclusion}

In this work, we report the discovery of shell features in three dwarf galaxies of stellar mass $\lesssim$10$^{8}~$M$_{\sun}$, VCC~1361, VCC~1447 and VCC~1668, located in the Virgo cluster. In the original VCC catalogue they are classified as dE morphology, and this is the first time such features have been detected in Virgo cluster dEs. One of them, VCC~1447, is fainter than the Local Group, Fornax dwarf galaxy, which also possesses a shell-like feature. However, compared to the feature detected in Fornax, the shell features reported here are much more pronounced. Their existence is not questionable as they were detected independently in the different bands provided by the NGVS.

We find that a major merger between two near equal mass, early-type dwarf galaxies can well explain the key properties of the observed shell systems in our sample.  The merger causes a large fraction of the stars to be disturbed, producing shells that contain a significant fraction of each galaxy's total light, and that can be located at large distances from their galaxy.  The general morphology and symmetry of the shells is well reproduced, as long as the merger is more major than a 1:2 mass ratio. The major merger scenario also causes the body of the galaxy to be elongated in the same direction as the interleaved shells, as observed. The small number of clear, well separated shells that are observed favors the hypothesis that a scenario in which the merger occurs recently, in the last few hundred Myrs. As such, the observed red NUV-optical color of all the sample, and lack of detectable color gradients or dust lanes, favors a dry merger scenario. The presence of merger features likely favors that the dwarfs are located in the cluster outskirts, and infalling into the cluster for the first time. 

We have studied three unique examples of dwarf galaxies with shell features and showed that major mergers are as the most likely origin. To get a more complete statistical picture of dwarf-dwarf mergers in the Virgo cluster, it will be important to perform a detailed investigation on all the other dwarf galaxies with candidate shell features in the NGVS.

\section*{Acknowledgments}
This work is supported by the French Agence Nationale de la Recherche (ANR) Grant Programme Blanc VIRAGE (ANR10-BLANC-0506-01).
We wish to express our gratitude to the CFHT personnel for its dedication and tremendous help with the MegaCam observations. RS acknowledges support from Brain Korea 21 Plus Program(21A20131500002) and the Doyak Grant(2014003730). EWP acknowledges support from the National Natural Science Foundation of China under Grant No. 11573002, and from the Strategic Priority Research Program, ``The Emergence of Cosmological Structures", of the Chinese Academy of Sciences, Grant No. XDB09000105. ET \& RG acknowledge the NSF grants AST- 1010039 and AST-1412504.
This research used the facilities of the Canadian Astronomy Data Centre operated by the National Research Council of Canada with the support of the Canadian Space Agency. The authors further acknowledge use of the NASA/IPAC Extragalactic Database (NED), which is operated by the Jet Propulsion Laboratory, California Institute of Technology, under contract with the National Aeronautics and Space Administration.


\end{document}